\documentclass[%
 aip,
 amsmath,amssymb,
 reprint,%
]{revtex4-2}

\usepackage{graphicx}
\usepackage{dcolumn}
\usepackage{bm}

\usepackage{color,ulem}
\usepackage{graphicx}
\usepackage{dcolumn}
\usepackage{bm}
\usepackage{color}

\usepackage[utf8]{inputenc}
\usepackage[T1]{fontenc}
\usepackage{mathptmx}
\usepackage{etoolbox}
\usepackage{soul}

\makeatletter
\def\@email#1#2{%
 \endgroup
 \patchcmd{\titleblock@produce}
  {\frontmatter@RRAPformat}
  {\frontmatter@RRAPformat{\produce@RRAP{*#1\href{mailto:#2}{#2}}}\frontmatter@RRAPformat}
  {}{}
}%
\makeatother
\begin{document}

\title{Modeling glasses from first principles using random structure sampling}

\author{Laszlo Wolf}                                                                                                     
\affiliation{Colorado School of Mines, Golden, CO 80401, USA}

\author{Andrew Novick}                                                                                                     
\affiliation{Colorado School of Mines, Golden, CO 80401, USA}

\author{Vladan Stevanovi\'{c}$^{\ast}$}
\email{vstevano@mines.edu}                                                                     
\affiliation{Colorado School of Mines, Golden, CO 80401, USA}

\date{\today}

%
\begin{abstract}
We present an approach to approximating static properties of glasses without experimental inputs rooted in the first-principles random structure sampling. In our approach, the glassy system is represented by a collection (composite) of periodic, small-cell (few 10s of atoms) local minima on the potential energy surface. These are obtained by generating a set of periodic structures with random lattice parameters and random atomic positions, which are then relaxed to their closest local minima on the potential energy surface using the first-principles methods. Using vitreous SiO$_2$ as an example, we illustrate and discuss how well various atomic and electronic structure properties calculated as averages over the set of such local minima reproduce experimental data. The practical benefit of our approach, which can be rigorously thought of as representing an infinitely quickly quenched liquid, is in that it transfers the computational burden to linearly scaling and easy to converge averages of properties computed on small-cell structures, rather than simulation cells with 100s if not 1000s of atoms while retaining a good overall predictive accuracy. Because of this it enables the future use of high-cost/high-accuracy electronic structure methods thereby bringing modeling of glasses and amorphous phases closer to the state of modeling of crystalline solids.  
\end{abstract}
%
\maketitle
%
\section{Introduction}\label{sec:introduction}
%
Despite of decades of research, glasses continue to puzzle the scientific community \cite{Parisi_Urbani_Zamponi_2020,Binder_Kob}. Both the nature of the glassy state, its formation and dynamics, as well as the relation of its properties to the underlying atomic structure continue to challenge our understanding of solid matter. Of course, much progress has been made towards developing both qualitative and quantitative understanding of the glassy state, but the challenges persist. In this paper we focus on atomically disordered glasses and the predictive methods to model their properties.  

Conventionally, glasses exhibiting atomic disorder are modeled using molecular dynamics (MD) simulations, either classical or ab-initio MD, and some version of the melt and quench procedure to construct representative structural models \cite{Raza_2015}. The accuracy of a structural model is typically assessed using a set of relevant properties, such as the mass density, various atomic correlation functions, etc., and the level of agreement between the calculated and the measured values  (see for example Refs.~\onlinecite{Pasquarello_PRB:1998,Massobrio_PRB:2009}). The physical picture underlying these modeling approaches is that of a continuous random network of atoms. In other words, the glassy state is thought of as a single configurational microstate adopted by the system. A practical disadvantage of these approaches is in that they typically require large simulation cells with 100s if not 1000s of atoms, which then presents challenges for computing electronic properties (optics, transport) using state-of-the-art, many-body electronic structure methods. Enabling the use of such methods would allow accurate calculations of various functional  properties of glasses, which will, in turn, allow for the better comparison with the results from spectroscopic measurements and investigation of the structure-property relationships.  

Herein we present an alternative method to modelling static properties of glasses in which a glassy state is approximated by a collection (composite) of microstates rather than a single one. The microstates are obtained using the first-principles random structure sampling in which a large set of small-cell, periodic structures with random cell parameters and atomic positions are generated and subsequently relaxed to the closest local minima on the potential energy surface using first-principles relaxation techniques. As we will show by taking glassy SiO$_2$ as a case example, averages over a set of these local minima reproduce measured properties including structure factors, pair distribution function and local coordinations, electronic density of states and bulk modulus, with sufficient accuracy allowing for a predictive computational methodology to be constructed. The somewhat larger, $\sim$9 \% overestimation of the mass density ($\sim$3 \% in linear dimensions) is analyzed in detail and the ways to improve the methodology, by restricting the averaging procedure to a subset of random structures, are discussed and outlined. The advantage of our approach is in the relatively small unit cells used to generate random structures, 24 atoms in the case of SiO$_2$ discussed here, which allows for more computationally demanding calculations to be conducted. Moreover, the method is easy to converge and is linearly scaling in the number of structures.   

The first-principles random structure sampling was first introduced by Pickard and Needs \cite{Pickard_JOPCM:2011} and was shown as an effective tool in searching for the ground-state structures, i.~e. the global minima on the potential energy surface, across various chemistries. In our prior work, we extended the methodology to polymorphism and  demonstrated how virtually all known crystalline phases in a variety of chemistries appear as the frequently occurring structures (local minima) in the random sampling \cite{Stevanovic_2016,Jones_2017,Jankousky_PRM:2023,Jankousky_JACS:2024}. Additionally, we showed how the mass density, pair distribution function, as well as configurational thermodynamics of  disordered systems, such as amorphous Si \cite{Jones_2020} and cation disordered ZnZrN$_2$, MgMoN$_2$ ternary nitrides \cite{Woods-Robinson_PRM:2022,Zakutayev_NS:2024}, can be accurately described by averaging over the ``small-cell'' local minima obtained using the first-principles random structure sampling. In this paper, we present an in-depth analysis of this approach and extend it from the structural, to electronic and mechanical properties.
%
\section{Methods}\label{sec:methods}
%
\subsection{First-principles random structure sampling} \label{ssec:random_structure_sampling}
%
The variant of the first-principles random structure sampling adopted in this work proceeds as follows:

\noindent {\it (i) Structure generation} starts with the choice of the number of atoms that are to be distributed inside the periodic simulation cell and the random choice of the cell parameters $(a, b, c, \alpha, \beta, \gamma)$. For practical reasons the range of allowed values is restricted to $0.8*scale \leq a,b,c \leq 1.6*scale$ and $60^{\circ} \leq \alpha, \beta, \gamma \leq 140^{\circ}$. The $scale$ of the structure is adjusted at the end so that the minimal distance between any two atoms is larger than some input value, typically close to the expected bond length.  In the next step atoms are distributed  inside the simulation cell. For this purpose a real-space grid is constructed with its vertices used as possible locations for the atoms. The grid is constructed using the planes corresponding to the minima and maxima of a plane wave $cos({\bf G} \cdot {\bf r})$ defined by the reciprocal lattice vector ${\bf G} = n_1{\bf g}_1 + n_2{\bf g}_2 + n_3{\bf g}_3$ with its crystal coordinates chosen randomly such that $4 \leq n_1, n_2, n_3 \leq 7$. Defining the grid that is not parallel to the unit vectors provides an additional degree of randomness and the bounds ensure that neither too dense nor too sparse sequence of discretized planes (superlattice) is constructed. The atoms are then distributed randomly over the vertices of the real-space grid with the constraint designed to prevent any two atoms to be too close to each other. This constraint is implemented by placing a gaussian of a given width (input parameter) on every occupied vertex. The next atom is then placed randomly over the set of points for which the value of the sum of gaussians centered on occupied vertices is below a certain threshold value. Lastly, in case of partially ionic systems such as SiO$_2$ studied here, where cation-anion coordination is preferred, different types of ions are distributed over different planes, that is, cations are distributed only on the planes corresponding to the minima and anions to the maxima of the plane wave. In this way different types of ions are interspersed  and the cation-anion coordination is favored already in the initial structures.

\noindent{\it (ii) Structure relaxations} are performed on each random structure using the standard density functional theory setup \cite{stevanovic_PRB:2012}. The electron-electron interactions are treated using the PBE exchange-correlation functional \cite{PBE_PRL:1996} and the projector augmented wave (PAW) method \cite{PAW_PRB:1994} as implemented in the VASP computer code \cite{Kresse_PRB_1999}. Plane-wave cutoff of 340 eV is used in all calculations, which is at least $\sim$20 \% higher than the recommended value for both Si and O pseudopotentials (normal PP for Si, and soft for O). Automatic generation of the $\Gamma$-centered ${\bf k}$-point grid is employed with the $R_k$ value of 20. The dependence of the results on the choice of the exchange-correlation functional is analyzed between the LDA \cite{LDA_PRB:1981}, PBEsol \cite{PBEsol_PRL:2008}, and PBE functionals. All degrees of freedom, including volume, cell shape and atomic positions, are relaxed using the conjugate gradient algorithm. Volume and cell-shape relaxations are restarted at least three times for the real-space grid to be re-created. Structural relaxations are considered converged when maximal force on any atom gets below 0.02 eV/{\AA}, total energy stops changing by more than $10^{-5}$ eV and the hydrostatic pressure drops below 0.5 kbar. 

In this work we generated a total of 3,000, 24-atom and additional 1,000, 18-atom, and  1,000, 27-atom random structures which were subsequently relaxed using the PBE functional. The dependence of the results on the exchange-correlation functional is analyzed by relaxing the first 1,000 structures with other functionals. The convergence of various quantities with respect to the number of random structures is discussed in Section~\ref{sec:results}.   
%
\subsection{Reciprocal space structure functions} \label{ssec:reciprocal_space_structure_functions}
%
In the case of multicomponent systems there is no unique way to define various structure functions. Throughout this work we use the definitions corresponding to the Faber-Ziman \cite{Faber_Ziman_1965} formalism consistent with the experimental data used for comparison. Here we provide only the necessary definitions. More detailed descriptions are provided in the Appendices of this paper, which were written by consulting Refs.~\onlinecite{Keen_2001,Waseda_1980}. Within the Faber-Ziman formalism, the total structure factor $S^{X}({\bf q})$  is related to the coherent scattering intensity per atom $I^{coh}_{a}({\bf q})$ as measured by X-ray diffraction experiments as:
\begin{equation}\label{eq:def_FZ_tot}
    S^{X}({\bf q}) \, = \, \frac{1}{\left< f(q)\right>^{2}} 
    \, I^{coh}_{a}({\bf q}) \, - \, \frac{\left< f^{2}(q)\right> - \left< f(q)\right>^{2}}{\left< f(q)\right>^{2}},
\end{equation}
where $\textbf{q}$ is the scattering vector and $q$ is its norm. $\langle f(q)\rangle = \sum_{\alpha} c_{\alpha}f_{\alpha}(q)$ is the average X-ray atomic form factor with $c_{\alpha}$ being the atomic fraction of the atoms of type $\alpha$ and $f_{\alpha}(q)$ is the atomic form factor of the same atom type. Correspondingly, $\langle f(q)^2\rangle$ is the average squared form factor. The  atom type resolved, or partial Faber-Ziman structure factors $S_{\alpha\beta}({\bf q})$ are defined as:
\begin{equation}\label{eq:FZ_part}
 \begin{gathered}
    S_{\alpha\beta}({\bf q}) \, = \\
    = \, \frac{1}{\sqrt{c_{\alpha}c_{\beta}}} \left[ \frac{1}{\sqrt{N_{\alpha}N_{\beta}}} \, \left( \sum_{j\in \alpha}^{N_{\alpha}}\sum_{k\in \beta}^{N_{\beta}} \, e^{-i{\bf q}{\bf r}_{jk}} \right) - \sqrt{N_{\alpha}N_{\beta}} \Delta_{{\bf q}0} \right] - \\
    - \frac{1}{c_{\beta} \Delta_{\alpha\beta}} + 1, \\ 
\end{gathered}
\end{equation}
where $N_{\alpha}$ and $N_{\beta}$ are the total number of $\alpha$ and $\beta$ type atoms, ${\bf r}_{jk}=\textbf{r}_{j}-\textbf{r}_{k}$ is the relative position vector between atom $j$ of the type $\alpha$ and atom $k$ of the type $\beta$, and $\Delta_{{\bf q}0}$ is the Kronecker delta that is introduced for the purpose of removing the forward scattering (${\bf q}=0$) term from the double sum. Note that the above definition implies that partial structure factors do not depend on the atomic form factors and are consequently also independent on the experimental measurement technique. The total structure factor on the other hand is dependent on the form factors. The total Faber-Ziman structure factor $S^{X}({\bf q})$ from Eq.~(\ref{eq:def_FZ_tot}) is obtained as a weighted sum of the partials through:
\begin{equation}\label{eq:total_S_sum}
S^{\mathrm{X}} ({\bf q}) = \sum_{\alpha}\sum_{\beta} \, c_{\alpha}c_{\beta} \frac{f_{\alpha}(q) f_{\beta}(q)}{\langle f(q) \rangle^2} S_{\alpha\beta} ({\bf q}),
 \end{equation}
where both sums run over all different atom types in the system. The total Faber-Ziman structure factor from neutron diffraction $S^{N}({\bf q})$ is obtained similarly through Eqs.~(\ref{eq:def_FZ_tot}) and (\ref{eq:total_S_sum}) by replacing the X-ray atomic form factors $f_{\alpha}(q)$ with the $q$-independent neutron coherent scattering lengths $b_{\alpha}$.

Throughout this work the above equations are employed to compute various structure functions of each (DFT relaxed) random structure. Periodicity of the random structures however, allows making the following simplifications. For every pair of atoms their relative position can be written as ${\bf r}_{jk} = ({\bf R} + {\bf u}_j) - ({\bf R}' + {\bf u}_k)$, with ${\bf R}$ and ${\bf R}'$ being the lattice vectors and  ${\bf u}_j$,${\bf u}_k$ positions of atoms inside the unit cell. The double sum in Eq.~(\ref{eq:FZ_part}) will then split into a double sum of $exp(-i{\bf q}({\bf R} - {\bf R}'))$ over ${\bf R}$ and ${\bf R}'$ multiplied by the double sum of  $exp(-i{\bf q}({\bf u}_j - {\bf u}_k))$, which goes only over the atoms of the type $\alpha$ and $\beta$ inside the unit cell. Because of periodicity, the first sum has non-zero value only if ${\bf q}$ is a reciprocal lattice vector. Hence, the Eq.~(\ref{eq:FZ_part}) can only be evaluated for  ${\bf q}$ corresponding to reciprocal lattice vectors with the double sum going over the atoms inside the simulation cell. Also, because we are interested in modeling glasses, which in most cases are isotropic and only exhibit $q$-dependence, and because when ensemble averaging various functions the relative orientations of their Brillouin zones is unknown, when computing $S_{\alpha\beta}({\bf q})$ from Eq.~(\ref{eq:FZ_part}) the exponential term is replaced by its spherical average $sin(qr_{jk})/qr_{jk}$.

\subsection{Real space structure functions} \label{subsec:real_space_structure_functions}
For isotropic, multicomponent systems the atom type resolved, or partial, pair distribution functions (partial PDFs) $g_{\alpha\beta}(r)$ are standardly defined as:
\begin{equation}\label{eq:pdf_0}
g_{\alpha\beta} (r)  = \frac{1}{4 \pi r^{2} n c_{\beta}} \left[ \frac{1}{N_{\alpha}} \left( \sum_{j\in \alpha}^{N_{\alpha}}\sum_{k\in \beta}^{N_{\beta}} \delta(r-r_{jk}) \right) - \Delta_{\alpha\beta}\delta(r) \right],
\end{equation}
with $n$ being the global concentration of atoms (number density), $r$ the distance variable, as before $r_{jk} = |{\bf r}_j - {\bf r}_k|$ the distance between atom $j$ of the type $\alpha$ and atom $k$ of the type $\beta$, $\Delta_{\alpha\beta}$ is the Kronecker delta, and $\delta(r)$ is the Dirac delta function. The second term in the brackets ensures that $g_{\alpha\alpha} (r=0) = 0$. Equivalent, but more practical (for programing) definition that is used in this work is:
\begin{equation}\label{eq:pdf}
g_{\alpha\beta} (r) \, = \, \left< \, \frac{1}{4\pi r^2 n c_{\beta}} \, \frac{dN_{\alpha\beta}(r)}{dr} \,\right>_{\alpha},
\end{equation}
where $dN_{\alpha\beta}(r)$ represents the number of $\beta$ atoms in a spherical region of the thickness $dr$ at the distance $r$ from a particular $\alpha$ atom, and the angle brackets $\langle\,\rangle_{\alpha}$ denote an average over all $\alpha$ atoms. It is assumed that in the above equation the $g_{\alpha\alpha} (r=0) = 0$ condition is strictly enforced. As defined in Eq.~(\ref{eq:pdf}), $g_{\alpha\beta} (r)$ measures the concentration of $\beta$ atoms at a distance $r$ from the $\alpha$ atoms relative to the global concentration of atom of type $\beta$ ($n c_{\beta}$). In disordered systems and at large distances these two concentrations are the same and $g_{\alpha\beta} (r)$ asymptotes to 1 for $r \rightarrow \infty$.

Having $g_{\alpha\beta} (r)$ allows calculations of the average coordination numbers of the atoms of the type $\alpha$ by the $\beta$ atoms in the following way:
 \begin{equation}\label{eq:coord}
\Delta N_{\alpha\beta} (r_1,r_2) \, = \, n c_{\beta} \, \int_{r_1}^{r_2} \, 4\pi r^2 \, g_{\alpha\beta} (r) \, dr,
\end{equation}
with $r_1$ and $r_2$ bounding the range of distances within which the coordination is evaluated. Another widely used real space structure function is the reduced PDF with its partials $D_{\alpha\beta}(r)$ definition being:
\begin{equation}\label{eq:D(r)}
D_{\alpha\beta}(r) \, = \, n 4 \pi  r  \left( \, g_{\alpha\beta}(r) \, - \, 1 \, \right).
\end{equation}

The $D_{\alpha\beta}(r)$ and the spherically averaged $S_{\alpha\beta}(q)$ from Eq.~(\ref{eq:FZ_part}) are related (in practice) through the sine Fourier transform in the following way:
\begin{equation}\label{eq:partial_FT}
    D_{\alpha\beta}(r) \, =  \, \frac{2}{\pi} \int_{0}^{q_{max}} q \, ( S_{\alpha\beta}(q)-1) \, sin(qr) \, M(q) \, dq,
\end{equation}
with $q_{max}$ representing the upper limit of the measured scattering vector, and $M(q)$ being a modification function introduced to minimize the effects of the truncation in the sine Fourier transform \cite{Lorch_1969}. A typical form of $M(q)$ that is also used in this work for comparisons with experimental results is:
\begin{equation}\label{eq:Lorch}
    M(q) = \frac{q_{max}}{q\pi}sin(\frac{q\pi}{q_{max}}).
\end{equation}
Similar to the reciprocal-space functions, the partial real-space functions from Eqs.~(\ref{eq:pdf}) and (\ref{eq:D(r)}) are independent on the scattering power of the involved atoms. The total (geometric) PDF, that measures relative concentration of any type of atom at the distance $r$ from any other atom can be obtained from the partials as:
\begin{equation}
g(r) \, = \, \sum_{\alpha}\sum_{\beta} \, c_\alpha c_{\beta} \, g_{\alpha\beta}(r),
\end{equation}
and similar for the total reduced PDF $D(r)$. However, this is often not what is reported from experiments. What is typically measured is the total structure factor (not the partials), which is then sine Fourier transformed to the real space. The total reduced PDF resulting from this procedure is then:
\begin{equation}\label{eq:D_FT}
\begin{gathered}
D^{X}(r) \, = \, \frac{2}{\pi} \, \int_0^{q_{max}} \, q \, (S^{X}(q)-1) \, sin(qr) \, M(q) \, dq \, = \\
= \, \frac{2}{\pi} \, \int_0^{q_{max}} \sum_{\alpha} \sum_{\beta} c_{\alpha}c_{\beta} \frac{f_{\alpha}(q)f_{\beta}(q)}{\langle f(q) \rangle^2} \, \times \\
\times \, q \, (S_{\alpha\beta}(q) - 1) \, sin(qr) \, M(q) \, dq.\\
\end{gathered}
\end{equation}
The $q$-dependence of the X-ray atomic form factors prevents taking the factor $c_{\alpha}c_{\beta}f_{\alpha}(q)f_{\beta}(q)/\langle f(q) \rangle^2$ in front of the integral and expressing the total $D(r)$ in terms of its partials $D_{\alpha\beta}(r)$. This can be done by adopting the Warren \textit{et al.} approximation \cite{Warren_1936}, where the form factors are written as $f_{\alpha}(q) = Z_{\alpha} \, h(q)$ with $Z_{\alpha}$ being the atomic number, and $h(q)$ an approximately universal, atom-independent function of $q$. In this way the fraction $f_{\alpha} f_{\beta}/\langle f \rangle^{2} \approx Z_{\alpha} Z_{\beta}/\langle Z \rangle^{2}$ and is $q$-independent. Such approximation is not needed for neutron diffraction as the scattering lengths $b_{\alpha}$ do not depend on $q$. Using the Warren approximation for the X-ray diffraction the total reduced PDF and the corresponding PDF from either the the X-ray or the neutron diffraction can be derived from Eq.~(\ref{eq:D_FT}) and expressed as the weighted sums of their partials as:
\begin{equation}\label{eq:pdf_tot_2}
\begin{gathered}
D^{X/N}(r) \, = \, \sum_{\alpha}\sum_{\beta} \, W^{X/N}_{\alpha\beta} \, D_{\alpha\beta}(r) \, ; \\
g^{X/N}(r) \, = \, \sum_{\alpha}\sum_{\beta} \, W^{X/N}_{\alpha\beta} \, g_{\alpha\beta}(r),
\end{gathered}
\end{equation}
where the weights $W^{X/N}_{\alpha\beta}$ are:
\begin{align}\label{eq:weights}
    W^{X}_{\alpha\beta} \, &= \frac{c_\alpha c_{\beta} Z_{\alpha} Z_{\beta}}{\langle Z \rangle^{2}}, & W^{N}_{\alpha\beta} \, &= \frac{c_\alpha c_{\beta} b_{\alpha} b_{\beta}}{\langle b \rangle^{2}}.
\end{align}
This distinction between the geometric $D(r)$ and $g(r)$ and their X-ray or neutron counterparts $D^{X/N}(r)$ and $g^{X/N}(r)$ is important when comparing calculated with measured PDFs. The comparisons presented in this work are always done using calculated quantities that correspond directly to what has been done in experiments.  
%
\subsection{Averaging properties}\label{ssec:ensemble_averaging}
%
As previously noted, we represent the glassy state by a collection of structures obtained using the random structure sampling.  In this approach properties of glasses are then obtained by averaging. Equations for averaging various structure functions follow directly from Eq.~(\ref{eq:def_FZ_tot}). Namely, if we adopt a picture of a glass as an composite of local environments corresponding to the relaxed random structures, then the total scattering intensity $I^{coh}_{a}(q)$ per atom is simply going to be the average of the individual scattering intensities. In this picture the possibility of multiple scattering of a single photon is neglected. The direct consequence of Eq.~(\ref{eq:def_FZ_tot}) is that the same averaging applies to the structure factor. Hence, the average total structure factor is computed as:
\begin{equation}\label{eq:S(Q)_ensemble_average}
S^{X/N} (q) \, = \, \sum_s \, P_s \, S_s^{X/N} (q) 
\end{equation}
where the sum goes over all structures from the random sampling, $P_s$ is the probability of the structure $s$ in the random sampling (its normalized frequency of occurrence), and $S_s^{X/N} (q)$ is the structure factor of the structure $s$ from either the X-ray or neutron diffraction. Since all random structures have the same chemical compositions the same averaging applies to the partial structure factors from Eq.~(\ref{eq:FZ_part}). Lastly, from Eqs.~(\ref{eq:D(r)}) and (\ref{eq:partial_FT}) the averaging rule for the PDF follows directly:
\begin{equation}\label{eq:g(r)_ensemble_average}
 g_{\alpha\beta}(r) \, = \, \sum_{s} \, \frac{n_{s}}{n} \, P_{s}  \, g_{\alpha\beta}^{(s)}(r),
\end{equation}
with $n_{s}$ standing for the concentration of all atoms of the structure $s$, $n=\sum_s \, P_s \, n_s$ is the average concentration, and $g_{\alpha\beta}^{(s)}(r)$ is the partial PDF of the structure $s$.  Lastly, average mass density is computed by dividing the mass of one unit cell with the average volume, as the volume and mass are experimental observables.
 
An approach for averaging other properties such as electronic density of states  and bulk modulus will be presented in Sections~\ref{ssec:el_dos} and \ref{ssec:bulk_mod}. Finally, it is important to note that averaging according to Eqs.~(\ref{eq:S(Q)_ensemble_average}) and (\ref{eq:g(r)_ensemble_average}) represent the high-temperature limit of the standard classical statistical mechanics ensemble averages that employs Boltzmann probabilities $P_{s} \sim e^{-\varepsilon_s/k_BT}$. The reasons to use the high-temperature limit rather than the finite temperature averages or some other choice of the probability distribution is discussed in Section~\ref{ssec:temp}.

\begin{figure*}
\includegraphics[width=\textwidth]{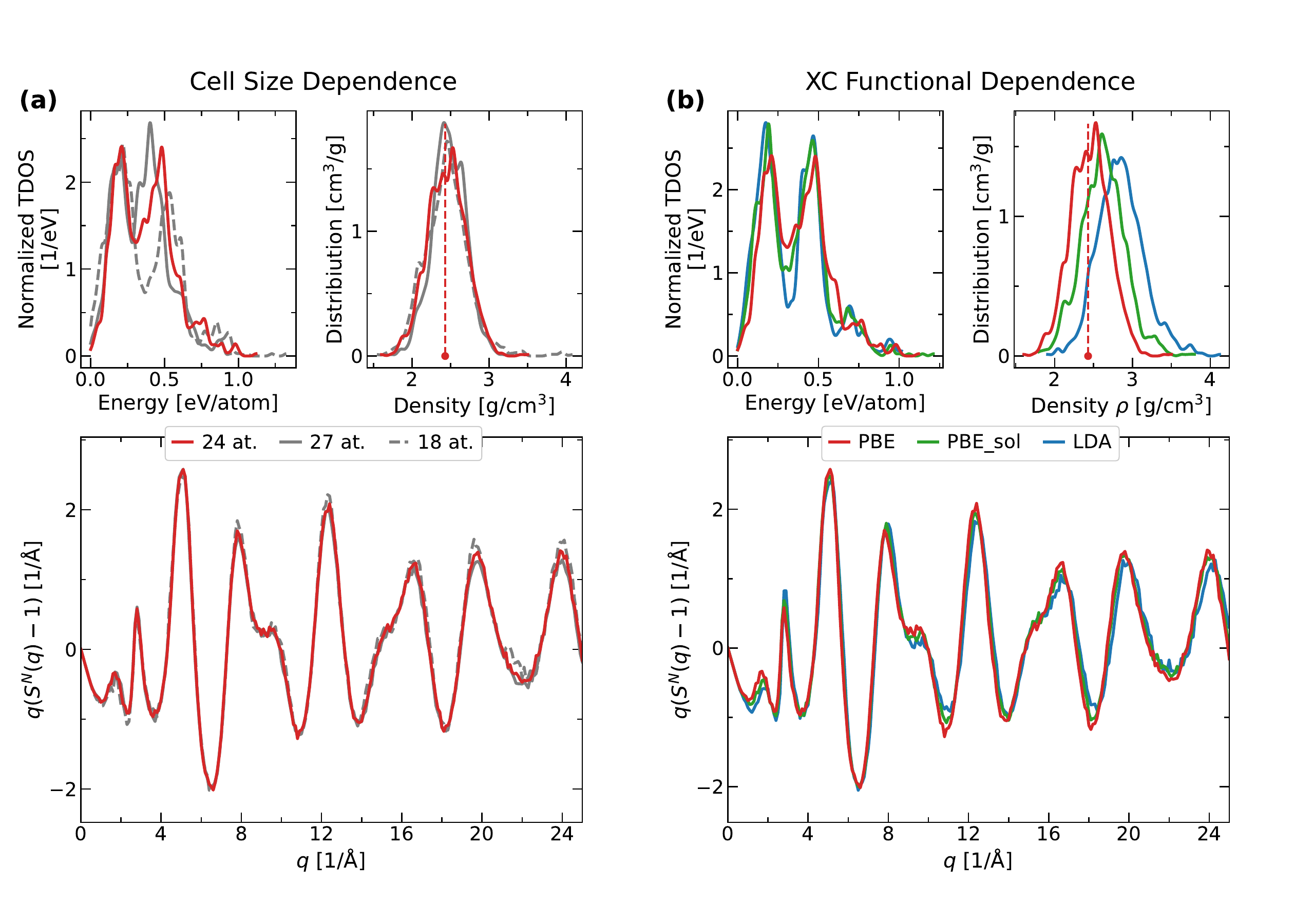}
\caption{\label{fig:XC_comparison} Dependence of the results on (a) the cell size, and (b) on the exchange-correlation functional used in calculations. Normalized Thermodynamic Density of States (TDOS), the mass density distribution over the random structures, and the average neutron structure factor are shown. Gaussian broadening of 0.015 eV and 0.035 g/cm$^{3}$ was used for the TDOS and the density distributions, respectively. Dashed vertical lines show the average density only for the PBE density distribution.}
\end{figure*}

%
\section{Results and discussion}\label{sec:results}
%
\subsection{Atomic structure -- Dependence on the cell size and the exchange-correlation functional}
%
We begin by analyzing the relaxed structures and their energy distribution, as well as the dependence of the results on the cell size, and the exchange-correlation functional used in relaxations. The comparison between different exchange-correlation functionals is performed on the first 1,000 random structures of the entire set of 3,000, 24-atom structures. These were relaxed using each of the following, LDA  \cite{LDA_PRB:1981}, PBE \cite{PBE_PRL:1996}, and PBEsol \cite{PBEsol_PRL:2008} functionals. The size dependence is analyzed at the PBE level only, by generating additional 1,000 structures with 18 atom cells, and 1,000 structures with 27 atom cells. These additional sizes are chosen because they are integer multiples of the number of atoms in one unit cell of the SiO$_2$ ground state, $\alpha$-quartz.

First, it is important to note that the vast majority of DFT relaxations, irrespective of the cell size and the functional, ended in the low symmetry local minima. As expected the fraction of more symmetric local minima decreases with size due to the exponential growth of the total number of local minima with the number of atoms\cite{Stillinger_PRE:1999}, the majority of which will have low symmetry. In case of 18-atom structures the PBE relaxations produce a total of 25 structures with space group numbers larger or equal to 10. Among these structures, there are 4 occurrences of the $\alpha$-quartz structure (space group \#154), 4 occurrences of a structure with s.g.\#41, three occurrences of a structure with s.g.\#15, two of a structure with s.g.\#155, and 12 others appearing only once. 

In the case of 24-atom cells, this number decreases to only one occurrence of a structure with s.g.\#15 and one with s.g.\#12. Extending to the full set of 3,000, 24-atom structures results in additional single occurrences of structures with s.g.\# 14, 71, and 148. All these can be treated as statistically insignificant in comparison to the overwhelming majority of the low-symmetry structures. For the 27-atom set, there are no structures with s.g.\# larger than 2 among 1,000 random structures. These results confirm the already well documented effectiveness of the random structure sampling as a structure prediction tool (global minimum search), especially for relatively small cells ($\sim$20 atoms or less).

However, the focus of this paper is not on the individual structure types, as the structure prediction would demand, but on the entire set of local minima including the low-symmetry ones and their utility in describing the glassy state.  Having 1,000 random structures is likely insufficient for the converged structure predictions to be attained, even for the 18-atom cells. Here, the convergence would imply finding all of the statistically significant (frequently occurring) local minima, and stabilizing their frequencies of occurrence as a function of the number of random structures \cite{Stevanovic_2016}.  This is however not the case if one is interested in representing glassy state. First, as will be discussed in details in Section~\ref{ssec:convergence}, both the averaged mass density and the structure factor are well converged even with 1,000 random structures. It is also evident from Fig.~\ref{fig:XC_comparison} that the average mass density in the high-temperature limit, as well as the average structure factor are both well-converged with respect to the cell size, and that 24-atom cells are sufficiently large for the converged results to be obtained.  The differences in the thermodynamic density of states (TDOS) seen in Fig.~\ref{fig:XC_comparison}(a) are to be expected, as the total number of states varies significantly for different number of atoms (exponential growth). In our previous study of amorphous silicon \cite{Jones_2020} we showed that the cell-size in the random sampling has to be larger than a certain threshold value (16 atoms in that case) for the sharp, crystalline-like features in the average X-ray diffraction to disappear. In the case of SiO$_2$, it appears that this condition is fulfilled even with 18-atom cells.

%
\begin{figure*}[t]
\includegraphics[width=0.85\textwidth]{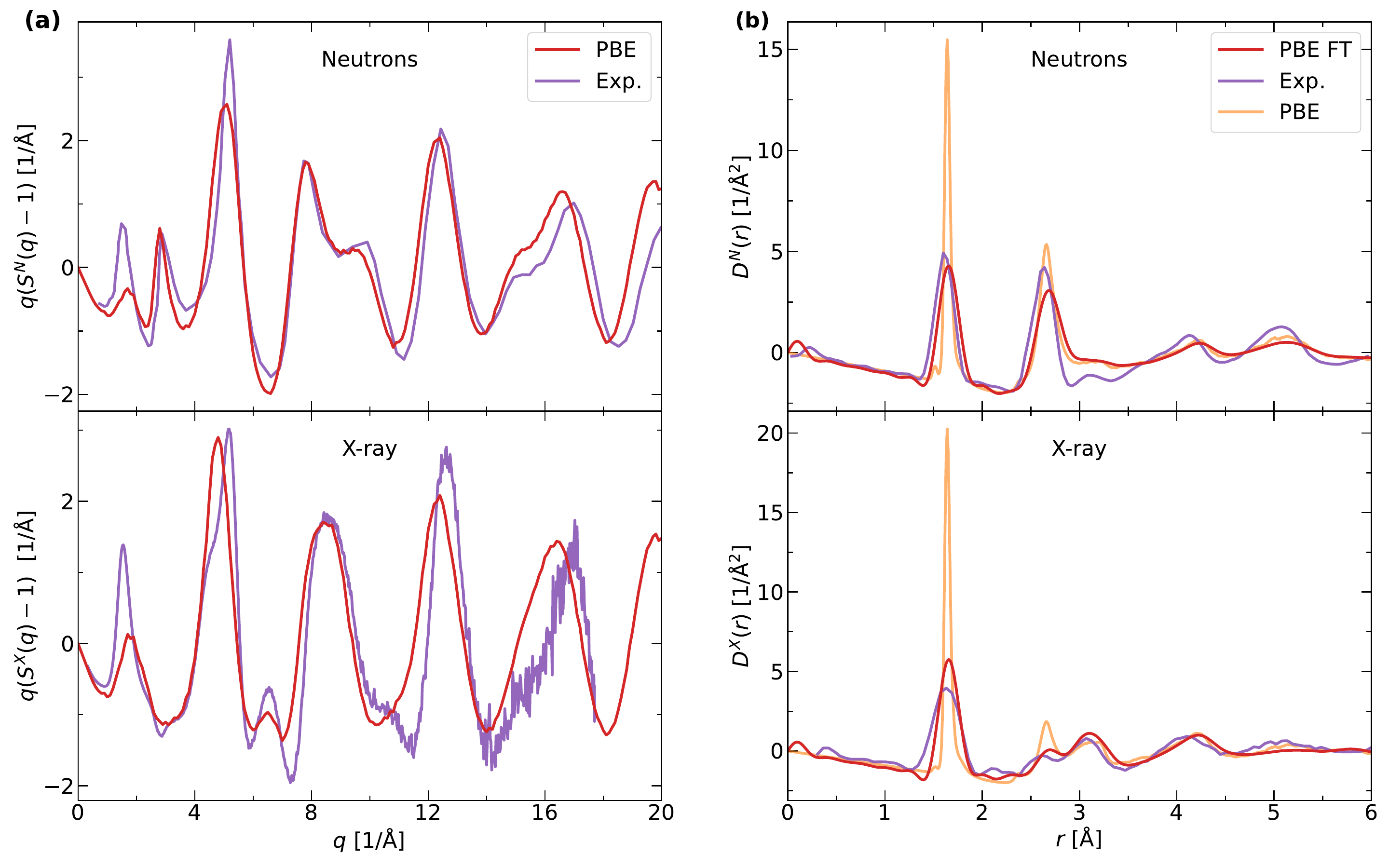}
\caption{\label{fig:PBE_vs_exp} Neutron and X-ray structure functions from computation (red) and experiment\cite{Susman_1991,Mei_2007} (purple). Panel (a) shows reciprocal space function $q(S^{N/X}(q)-1)$ and panel (b) the reduced pair distribution function $D^{N/X}(r)$ calculated both directly in real space (orange curve) and as the sine Fourier transform (red), both averaged over 3000 random structures (see text for details).}
\end{figure*}
%
Different exchange-correlation functionals also seem to provide very similar descriptions of the potential energy surface. The TDOSs resulting from relaxations of the same 1,000 initial random structures with different functionals presented in Fig.~\ref{fig:XC_comparison}(b) display similar behavior as a function of energy for all sets with a two-peak shape centered around 0.2 and 0.5 eV above their respective lowest energy structures. While peak intensities vary somewhat, the similarity between curves shows that different exchange-correlation functionals will provide an overall similar distributions of structures over energies.  Similarities between different exchange-correlation functionals are also evident from Fig.~\ref{fig:XC_comparison}(b) showing distribution of mass densities. All distributions present an approximately Gaussian shape, with the main difference in the average value of the density. As expected, the LDA functional produces on average higher density structures while PBE yields lower density structures. The PBEsol is in between. 

Lastly, the ensemble-averaged structure factors shown in Fig.~\ref{fig:XC_comparison} are in good agreement, both between different sizes at the PBE level of theory and between different functionals for a fixed 24-atom cell size. In the remainder of the text we will be using the PBE functional with random sampling done on 24-atom cells.
%
\subsection{Atomic structure -- Comparison with experiments}
We first discuss the mass density. The average value calculated with the PBE functional is 2.4 g/cm$^3$ using either 1,000 or 3,000 structures, while PBEsol and LDA give 2.6 and 2.8 g/cm$^3$, respectively (1,000 structures). The errors relative to experimental result ($\sim$2.2 g/cm$^3$ from Ref.~\onlinecite{Susman_1991}) are: 9 \% (PBE),$\sim$18 \% (PBEsol), and $\sim$28 \% (LDA). This implies errors in linear dimensions of $\sim$3 \% for PBE,  $\sim$6 \% for PBEsol, $\sim$10 \% for LDA. The 3\%  PBE error in linear dimensions is slightly higher than the typical errors of 1-2 \% that DFT methods make in reproducing lattice constants of crystals\cite{Zhang_NJP:2018}. Furthermore, this error in linear dimensions is opposite to what would be expected for PBE functional. Namely, PBE usually produces lattice constants of crystals that are larger than the experiments \cite{Zhang_NJP:2018} while here it seems that all functionals produce overall shorter linear dimensions on average than experiments. Comparison of the structure factors and the PDFs, which we do next will give us better insight into the origins of the somewhat larger than expected errors in the density and linear dimensions. 

The comparison between the ensemble-averaged structure factors $S(q)$ in the high-temperature limit obtained using the PBE exchange correlation functional and experimental ones is presented in Fig.~\ref{fig:PBE_vs_exp}(a) for both neutron and X-ray scattering. More precisely, the comparison is done on the $q(S(q)-1)$ rather than the $S(q)$ because the  former offers better insight into the agreement between theory and experiment, in particular for large $q$ values. In addition, the real-space reduced PDF $D(r)$ shown in Fig.~\ref{fig:PBE_vs_exp}(b) is obtained by the sine Fourier transform of $q(S(q)-1)$. Measured positions and intensities of all peaks in $q(S(q)-1)$ are in excellent agreement, with the exception of the first peak at $q \approx 1.5$ 1/{\AA} whose intensity is noticeably lower in our calculations than in measured $S(q)$. A general small shift of the calculated $S(q)$ to lower $q$-values can also be observed, which can be associated to the differences in the interatomic distances between the relaxed structures and the measurements. Importantly, the overall agreement with the experimental results is on par with previously reported first-principles molecular dynamics simulations of glassy SiO$_2$ \cite{Giacomazzi_PRB:2009}.

In Fig.~\ref{fig:PBE_vs_exp}(b), in orange the reduced PDF $D(r)$ computed in real space from Eqs.~(\ref{eq:pdf}) and (\ref{eq:D(r)}) are presented together with $D(r)$ obtained by the sine Fourier transform from the corresponding functions of the left panel, as would be done in experiments. Peak positions and overall shape are again in good agreement with experimental results. The difference in sharpness between the orange and red curves showcases the broadening effect of the Fourier transform. One can also note the better agreement for neutrons scattering of the FT data with experiment where for X-ray scattering our data is consistently sharper. Such differences can be associated with the lack of phonons in our description, as the X-rays measurements are typically done at room temperature while the neutron scattering is done at low temperatures. 

\begin{figure}
\includegraphics[width=\linewidth]{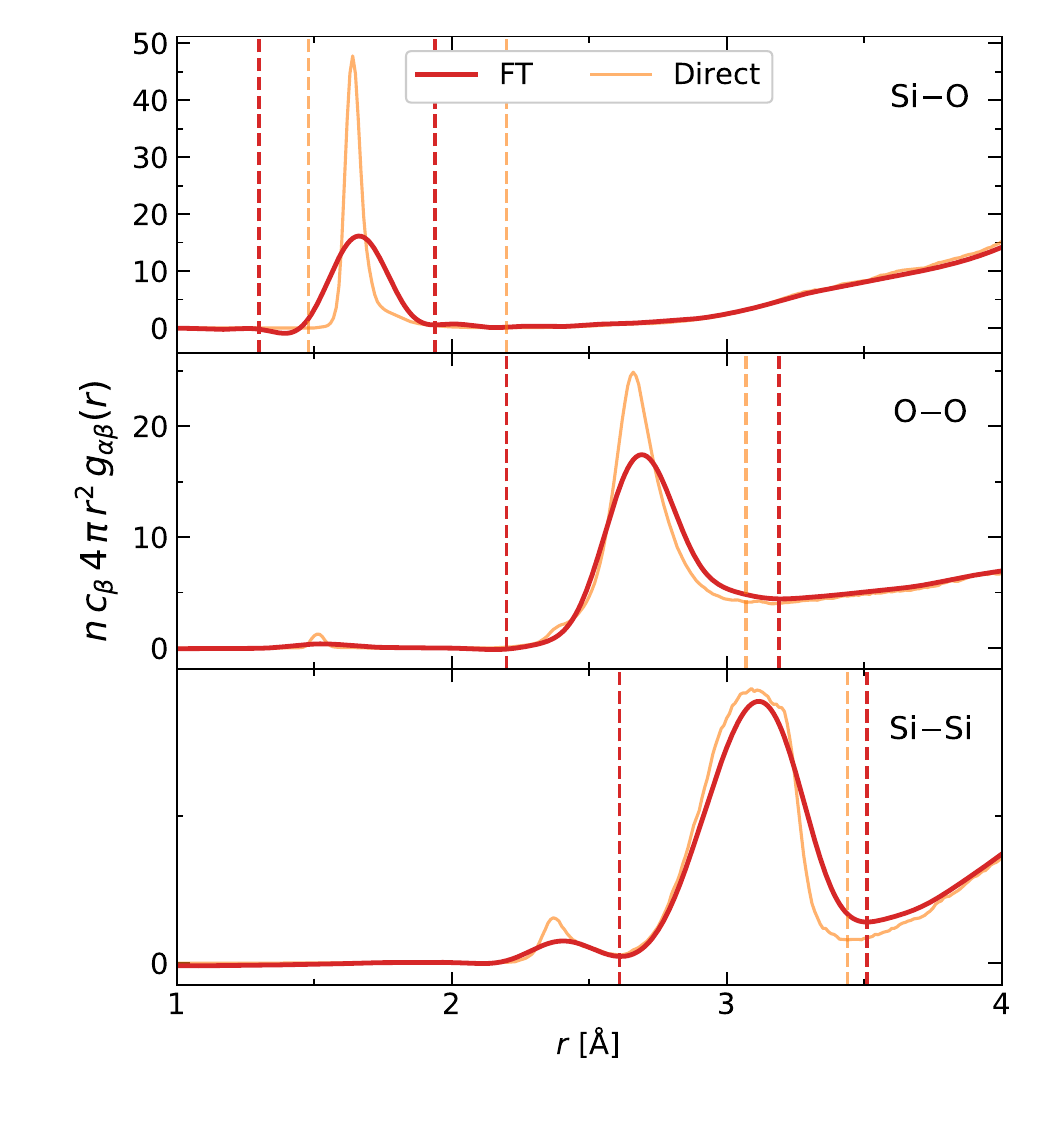}
\caption{\label{fig:coord_no} Atom-pair specific real space functions from which coordination numbers from Table~\ref{table:coord_no} are obtained using Eq.~(\ref{eq:coord}). Dashed vertical lines show integration bounds identifying the peaks, right bound is identified as the next local minimum.}
\end{figure}

A more quantitative real-space comparison is done by evaluating various coordination numbers. The procedure follows what would typically be done in experiments, that is, integration of of the $n c_{\beta} 4\pi r^2 g_{\alpha\beta}(r)$ within the ranges that are defined by the peak of interest in the corresponding partial PDF $g_{\alpha\beta}(r)$.  The start and the end of different peaks are in our work defined as corresponding to the local minima closest to the peak maximum as illustrated in Fig.~\ref{fig:coord_no} for both directly calculated and sine Fourier transformed functions by the vertical dashed lines. The computed coordination numbers from the first peaks of each partial  $g_{\alpha\beta}(r)$ are provided in Table~\ref{table:coord_no} together with those reported from experiments. Again, the overall agreement is rather good both in the peak positions and coordination numbers. All our data is well within the expected values for the peak positions and within the experimental error bars for the coordination numbers except for the O-O coordination number ($\Delta N_\mathrm{O-O}$) which is slightly overestimated. 

A significant portion of the deviations in the coordination numbers could be attributed to the numeric inaccuracies when  defining the domain of integration. It is evident from Fig.~\ref{fig:coord_no} that the definition of various peaks could be very different between the directly calculated $n c_{\beta} 4\pi r^2 g_{\alpha\beta}(r)$ curves and those obtained by the Frourier transforms. This is particularly true for peaks after the first Si-O peak. Hence, the effects of Fourier transform need to be taken into account when comparing calculated values to ones reported from experiments.  
%
\begin{table*}
\caption{\label{table:coord_no} Mass density, positions of the first peaks ($r_{\alpha - \beta}$) in the averaged partial PDFs $g_{\alpha\beta}(r)$ and the corresponding coordination numbers ($\Delta$N$_{\alpha - \beta}$). Comparison of the data from: direct, real-space analysis of the random structures (Calc. direct), data from partial structure factors calculated by Fourier transforming $q(S(q)-1)$ (Calc. FT), and  experiments.}
\begin{ruledtabular}
\begin{tabular}{cccccccc}
  & Density         & \multicolumn{2}{c}{Si-O}                                          & \multicolumn{2}{c}{O-O}                                        & \multicolumn{2}{c}{Si-Si}\\
  &   [g/cm$^3$] & $r_\textrm{Si-O}$ [{\AA}] & $\Delta N_\textrm{Si-O}$  & $r_\textrm{O-O}$ [{\AA}] & $\Delta N_\textrm{O-O}$ & $r_\textrm{Si-Si}$ [{\AA}] & $\Delta N_\textrm{Si-Si}$ \\
\hline
  &                     &                                &                                                 &                               &                                              &                                &                                               \\
Calc. direct & 2.43 & 1.64 & 3.93 & 2.66 & 6.47 & 3.08 & 3.79 \\
Calc. FT     & 2.43 & 1.66 & 3.76 & 2.69 & 7.10 & 3.12 & 3.91 \\
  &                     &                                &                                                 &                               &                                              &                                &                                               \\
Exp.1\footnote{Ref.~\onlinecite{Mei_PRB_2008}} & 2.20 & 1.60 $\pm$ 0.01 & 3.89 $\pm$ 0.20 & 2.62 $\pm$ 0.01 & 5.99 $\pm$ 0.30 & 3.08 $\pm$ 0.01 & 4.06 $\pm$ 0.20\\
Exp.2\footnote{Ref.~\onlinecite{Grimley_JONCM:1990}} & 2.20 & 1.608 $\pm$ 0.004 & 3.85 $\pm$ 0.16 & 2.626 $\pm$ 0.006 & 5.94 $\pm$ 0.20 \\
\end{tabular}
\end{ruledtabular}
\end{table*}
%

\begin{figure}
\includegraphics[width=\linewidth]{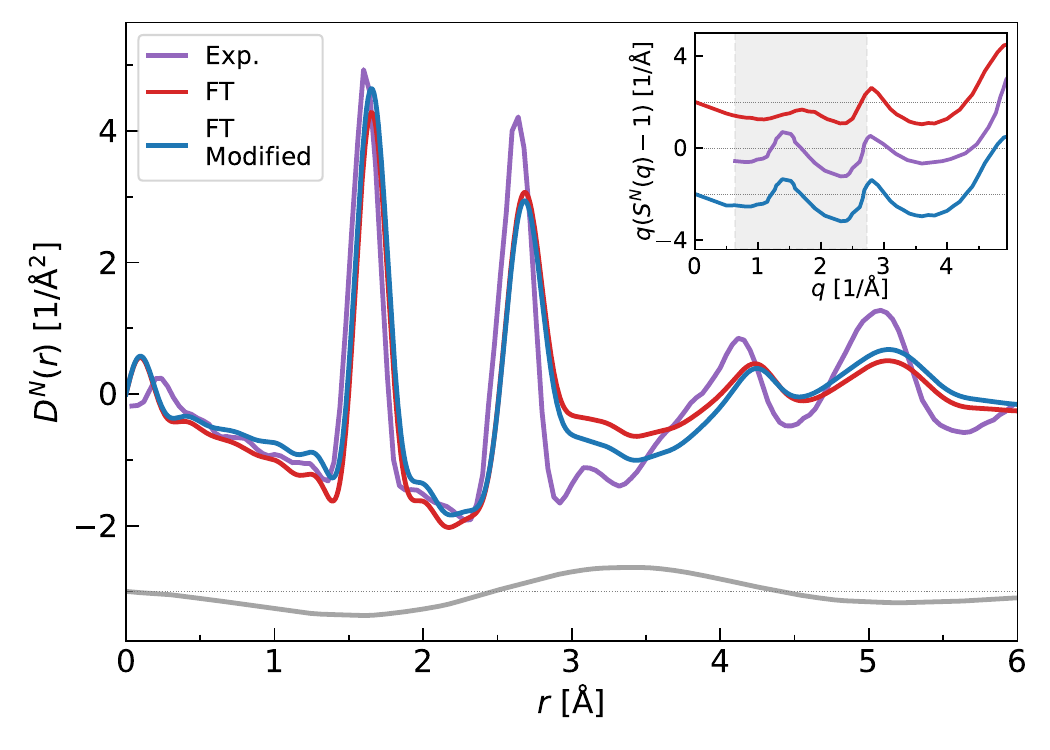}
\caption{\label{fig:modified} Reduced PDFs from sine Fourier transformation of both modified (blue) and unmodified (red) neutron diffraction structure factor and experiments\cite{Susman_1991} (purple). Difference between the red and blue curves is shown with an offset in grey. Inset shows the reciprocal space structure factors from which the reduced PDFs are obtained (vertical offset for clarity). The modified structure factor is obtained by replacing first peak (shaded region in the inset) with the experimental one.}
\end{figure}

Lastly, it is important to note that some of the differences in how the atoms are distributed in space, in particular related to the $r$-values larger than the size of the first few coordination shells can be traced back to the inaccuracy in reproducing the first diffraction peak in both $S^{N}(q)$ and $S^{X}(q)$  at $q \approx 1.5$ 1/{\AA}. While the origins of this diffraction peak are still being discussed, it has recently been linked with medium to large inter-atomic distances \cite{Biwas_PRB_2024}. Fig.~\ref{fig:modified} illustrates the effect of the first diffraction peak on both shorter and longer range atomic correlations. To elucidate the effect we replaced the first peak in the calculated $S^{N}(q)$ with the experimental curve and perform the sine Fourier transform. Fig.~\ref{fig:modified} shows the experimental $D(r)$, fully calculated $D(r)$, and the ``modified'' one obtained from the structure factor with the experimental first peak. Structure factors are shown in the inset. 

The short-range ($r \leq 6$ {\AA}) differences between the computed and modified $D(r)$ shown in the bottom of the figure, albeit small, are in the direction of bringing calculated values closer to the experimental ones. The long range effects are more visible from the initial slope of $D(r)$ before the first peak. Namely, the slope is directly proportional to the total concentration of atoms $n$ (number density). It is evident from the figure that the initial slopes of both experimental and modified $D(r)$ are less steep than that of the calculated $D(r)$ implying lower concentration. Consequently, mass density will be lower as well and is calculated to be below 2 g/cm$^3$ for the modified curve. Medium to long range differences that contribute to the first diffraction peak are most likely related to the ring structures (voids) with dimensions larger than ~3-4 {\AA} that are known to exist in the glassy SiO$_2$ \cite{Kono_NC:2022}. As shown in the next section, such voids can be observed in our structures only their number is likely underestimated if the entire set of structures is considered and thus contributing to said differences. The lack of vibrations and associated effects, such as the zero point motion and thermal expansion, could be among other possible contributors to the higher density in our model \cite{Zhang_NJP:2018}. However, it is important to note that the $\sim$9 \% overestimation of the density ($\sim$3 \% in linear dimansions) are still relatively small. Additionally, the nearsightedness of electronic matter \cite{prodan_PNAS:2005} would imply little influence on the long(er) range atomic correlations on functional properties, which we show in this paper.
%
%
\subsection{Emerging physical picture}\label{ssec:temp}
%
Representing the glassy state as the collection of small-cell local minima implies that the single microstate that is realized over the course of the glass transition can be approximated by a collection of local environments, each corresponding to a small-cell, periodic local minimum. This representation should not be confused with the crystallite hypothesis in which the glassy state is described as spatially local regions of crystalline material \cite{porai_JNCS:1990},  as most of the local minima obtained via the first-principles  random structure sampling have no symmetry at all. Hence, the emerging physical picture is that of a composite of local environments which correspond to the relatively small-cell periodic local minima on the potential energy surface. As having structures above a certain cell size is required for this approach to work, it is likely that the vast majority of these local minima will have little symmetry, if at all.

Second, computing various properties as averages over those local minima as explained in Section\ref{ssec:ensemble_averaging} is equivalent to the ensemble averages in the high-temperature limit. The implication is that our approach assumes full ergodicity over the entire set of the local minima obtained via the random structure sampling. This should not be mistaken for the full ergodicity over all local minima, as the ``unreasonable'' structures such as those having all Si atoms on one side and all O atoms on the other, are not part of our set. As the kinetic energy is not taken into account, the emerging physical picture is that of an instantaneously frozen, or infinitely quickly quenched liquid state. As a consequence, our approach, as described so far, is most appropriate for glasses that do not appreciably differ from their liquid state. In case significant changes in the structure occur during the glass transition, a possible solution would be to restrict the ergodicity so only the right subset of local minima is included in the averaging. How to properly take into account structural changes that would occur during the quenching process by restricting averaging to a subset of the random structures is an important avenue for further improvement of the presented methodology. 

To illustrate this point, we investigate the impact of the temperature parameter, which offers the simplest way of favoring certain subset of local minima (the low energy ones), on the final results. Before we do that it is important to note that in this context temperature should be used only as a parameter that influences the average values. This is mainly due to the underestimation of the total TDOS in our approach because of the already noted exponential increase in the number of local minima with the number of atoms. Not having correct total number of states implies underestimation of the configurational entropy and prevents rigorous formulation of the thermodynamic limit and absolute temperature. In case of alloys (lattice disorder) the high-temperature limit for the configurational entropy is known and the appropriate corrections have been devised to allow proper temperature dependencies to be evaluated \cite{Novick_2023}. However, this shortcoming does not prevent us though from investigating the effect of temperature, as a parameter, on the averaging. 
%
\begin{figure}
\includegraphics[width=\linewidth]{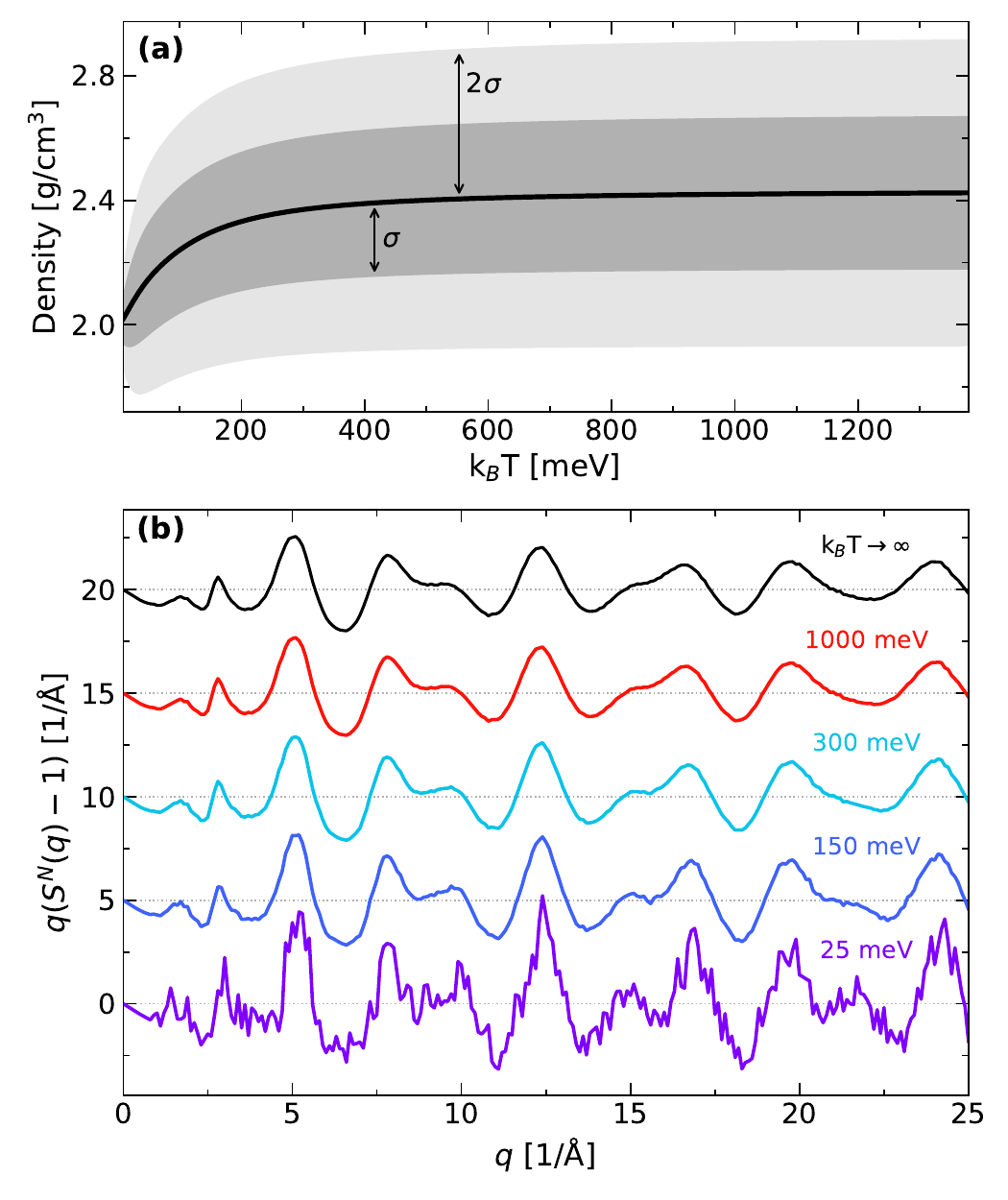}
\caption{\label{fig:temperature_Sq_rhodistrib} (a) The ensemble-averaged mass density and (b) neutron scattering structure factors as a function of ($k_{B}\mathrm{T}$). Shaded regions in (a) show one and two standard deviation intervals. The vertical offset between curves in panel (b) is for clarity.} 
\end{figure}
%

%
\begin{figure}
\includegraphics[width=\linewidth]{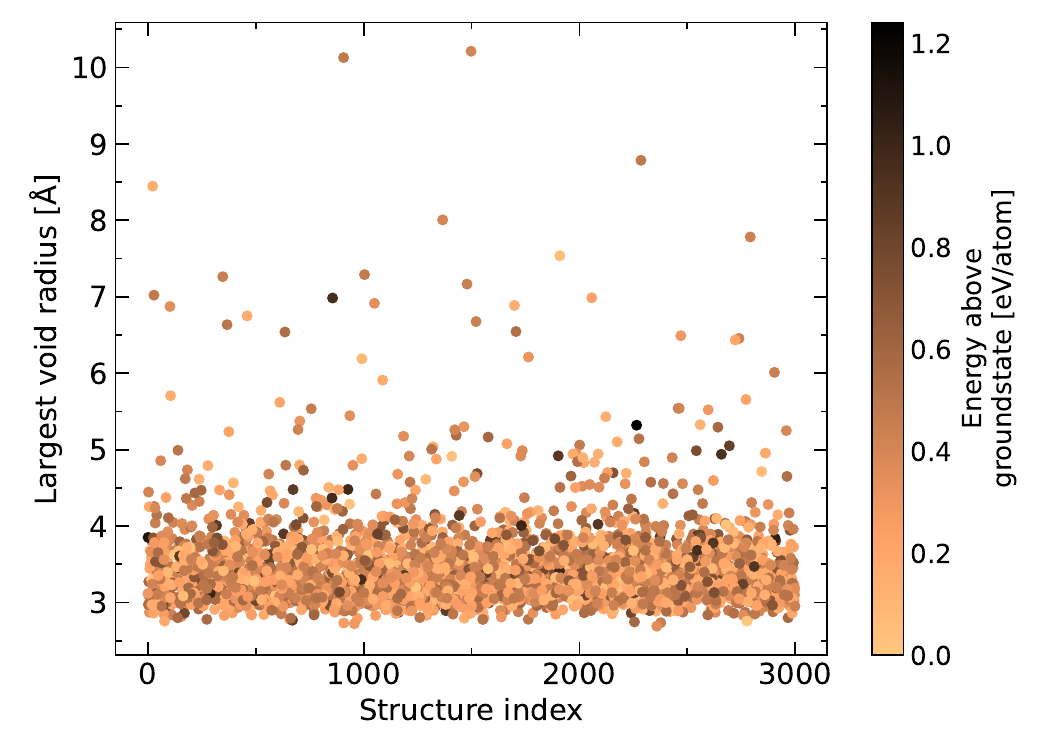}
\caption{\label{fig:voids} Scatter plot of the radii of the largest void in each of the $\sim$3,000 structures obtained using the PBE functional. Color gradient shows energy measured from the lowest energy structure. The method to identify the voids follows Kono {\it et al.}\cite{Kono_NC:2022}} 
\end{figure}
%

Fig.~\ref{fig:temperature_Sq_rhodistrib} shows the temperature dependence of the average mass density (upper panel) and $q(S^{N}(q)-1)$ (lower panel) with averaging performed using the total energy per formula unit inside the Boltzmann factor. It is clear from the figure that mass density increases with temperature, going from about 2 g/cm$^3$ at low temperature to the 2.4 g/cm$^3$ in the high-temperature limit. The experimental value of $\sim$2.2  g/cm$^3$ could be obtained by averaging with the temperature of $\sim$100 meV. The $q(S^{N}(q)-1)$ curves do not exhibit appreciable temperature dependence. As expected the curves are smoother as the temperature increases because of the larger number of structures contributing to the average.  Second effect is the decrease in the intensity of the first peak with increasing temperature, which is another evidence of the previously discussed connections of the first diffraction peak with mass density.

%
\begin{figure*}
\includegraphics[width=0.7\linewidth]{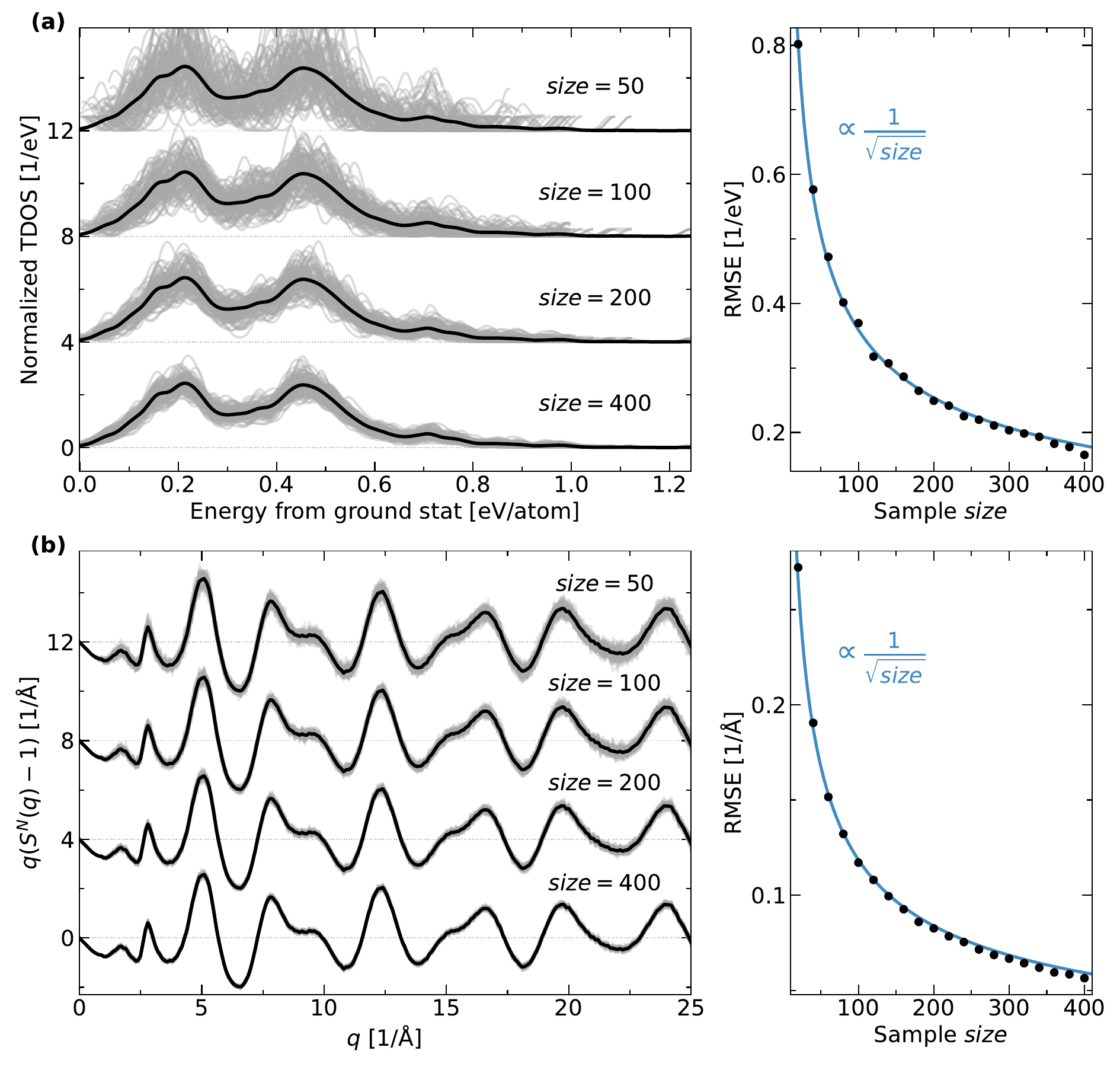}
\caption{\label{fig:TDOS_Sq_convergence}(a) Variations of the normalized TDOS and (b) neutron scattering structure factor, obtained from samples of different \textit{size} (number of structures in the sample). Left panels have vertical offset for clarity, solid black line shows the result using all 3,000 structures; grey curves show the results obtained from 100 samples with the indicated \textit{size}. Each sample is built by randomly picking structures from the 3,000 structure pool. Each point in the RMSE scatter plots is obtained as an average over the 100 samples of that \textit{size}.} 
\end{figure*}

Based on the temperature dependence of the results from Fig.~\ref{fig:temperature_Sq_rhodistrib} one could formulate a more pragmatic approach to modeling glasses in which the averaging temperature is chosen so to reproduce some measured quantity and is then used for predicting values of other relevant quantities. This is similar to the concept of microscopic fictive temperature introduced and discussed previously \cite{Mauro_JACS:2009,Mauro_JNCS:2009}. While certainly appealing, the fictive temperature approach has two major shortcomings. First, as discussed at length in the literature the temperature values that correspond to averages of different properties may be fairly different leading to the distribution of fictive temperatures, which then conflicts with our goal to have predictive methodology. A more fundamental criticism is related to using equilibrium theory to describe non-equilibrium phenomena \cite{Mauro_JACS:2009,Mauro_JNCS:2009}. 

While not exactly the same, our approach may be seen as a variation of the fictive temperature idea. The main difference is the assumed physical picture of the collection of local environments that are representing a single microstate of a glass rather than an ensemble of its microstates. Also, our averages are always done in the high-temperature limit exactly to remove the variability of temperatures in the averages, and lastly, our description is best aligned with the assumption of an infinitely quickly quenched liquid for which it was argued the concept of fictive temperature may be appropriate \cite{Mauro_JACS:2009}. In summary, our main argument is that the approach to modeling glasses presented in this paper, while certainly approximate, offers practical advantages and predictive power in modeling static properties of glasses.

To further elucidate previous points related to the overestimation of the mass density and restriction of the averaging to a subset of structures as a possible solution, we show in Fig.~\ref{fig:voids} the distribution of the radii of the largest void in each structure. It is clear that the cell size of 24 atoms does not preclude voids of ~3-4 {\AA} of appearing in our structures. The  method for the identifying voids follows directly Kono {\it et al.}\cite{Kono_NC:2022}. The resulting plot shows a varied distribution of such radius, ranging from ~3 {\AA} to ~10 {\AA} with most of the points concentrated in the 3-4 {\AA} range. This data clearly shows that the size of our cells isn't preventing the appearance of larger features and that a number of low-energy structures do contain them as also evident from Fig.~\ref{fig:temperature_Sq_rhodistrib}(a). However, the distribution of such features over the entire set of structures is underestimated. We would like to stress that even when considering the entire set of structures without any restrictions the $\sim$9\% overestimation of the density  (3\% in linear dimensions) still allows for predictive methodology to be formulated as illustrated by the agreement of previous and upcoming presented results with experimental data.

%
\subsection{Practical aspects -- How many structures is enough?}\label{ssec:convergence}
%
Before discussing property calculations, it is important to analyze the convergence of the results with respect to the number of random structures. Having a minimal set of random structures that produces the desired accuracy in property predictions is necessary as many of the properties may still be difficult to calculate for thousands of structures, even at the DFT level.  Fig.~\ref{fig:TDOS_Sq_convergence} shows the dependence of the normalized TDOS and $q(S^{N}(q)-1)$  on the number of random structures used in averaging. The averages over the entire set of 3,000 structures computed with the PBE functional are used as a reference (black curves). The convergence test is then performed by choosing randomly smaller number of structures (50, 100, 200, and 400) to average over. For each size, 100 random samples are created and the properties computed for each of them are shown as grey curves in Fig.~\ref{fig:TDOS_Sq_convergence}. The corresponding root-mean-square-errors (RMSE), averaged over the 100 samples, are also shown in the right panels. As expected from the central limit theorem the RMSE broadly follows the $1/\sqrt{\mathrm{size}}$ dependence on the number of structures in the sample\cite{Novick_2023,CLT_1,CLT_2}.

As evident from Fig.~\ref{fig:TDOS_Sq_convergence} many more structures are needed to converge the TDOS as opposed to $q(S^{N}(q)-1)$. This is again expected, because describing the energy distribution of local minima requires accumulating enough statistics. On the other hand, if the cell-size is such that many local minima already contain enough of the information about the atom-atom correlations, than less structures would be needed for the convergence of the structure factor to be  achieved. Indeed, the grey curves are much closer to the black in Fig.~\ref{fig:TDOS_Sq_convergence}(b), and, depending on the desired accuracy, as few as $\sim$100 random structures may be sufficient to achieve converged structure factor. This is a very important result as it allows to represent the glassy SiO$_2$ with only $\sim$100, 24-atom random structures, which is a small enough number for the more advanced electronic structure calculations to be performed. While doing advanced electronic structure calculations is out of the scope of this work, we will test the accuracy of the property calculations at the DFT level in the next two sections. The properties of interest include the electronic density of states as well as the bulk modulus of SiO$_2$. 
%
\subsection{Electronic properties -- Density of states}\label{ssec:el_dos}
%
We first analyze the electronic density of states as calculated by averaging the PBE electronic DOSs of the random structures. Averaging requires aligning the electronic energies between different structures, which is done in the following way. The representation of the glass as a composite of the local environments implies the unique value of the Fermi energy for the entire sample, further implying the alignment of the electronic energies of each random structure to the common Fermi energy. For each structure the Fermi energy is evaluated so to have the equilibrium between the holes and electrons established as defined by the electronic DOS and the Fermi-Dirac distribution function in the low-temperature limit (1 \% of the band gap). For this purpose electronic DOS of each random structure is computed non-selfconsistently on a denser k-point grid having a total of $\sim$250 k-points (number of atoms $\times$ number of k-points = 6000). Each electronic DOS is broadened using the Gaussian smearing of 0.01 eV.
%
\begin{figure}
\includegraphics[width=\linewidth]{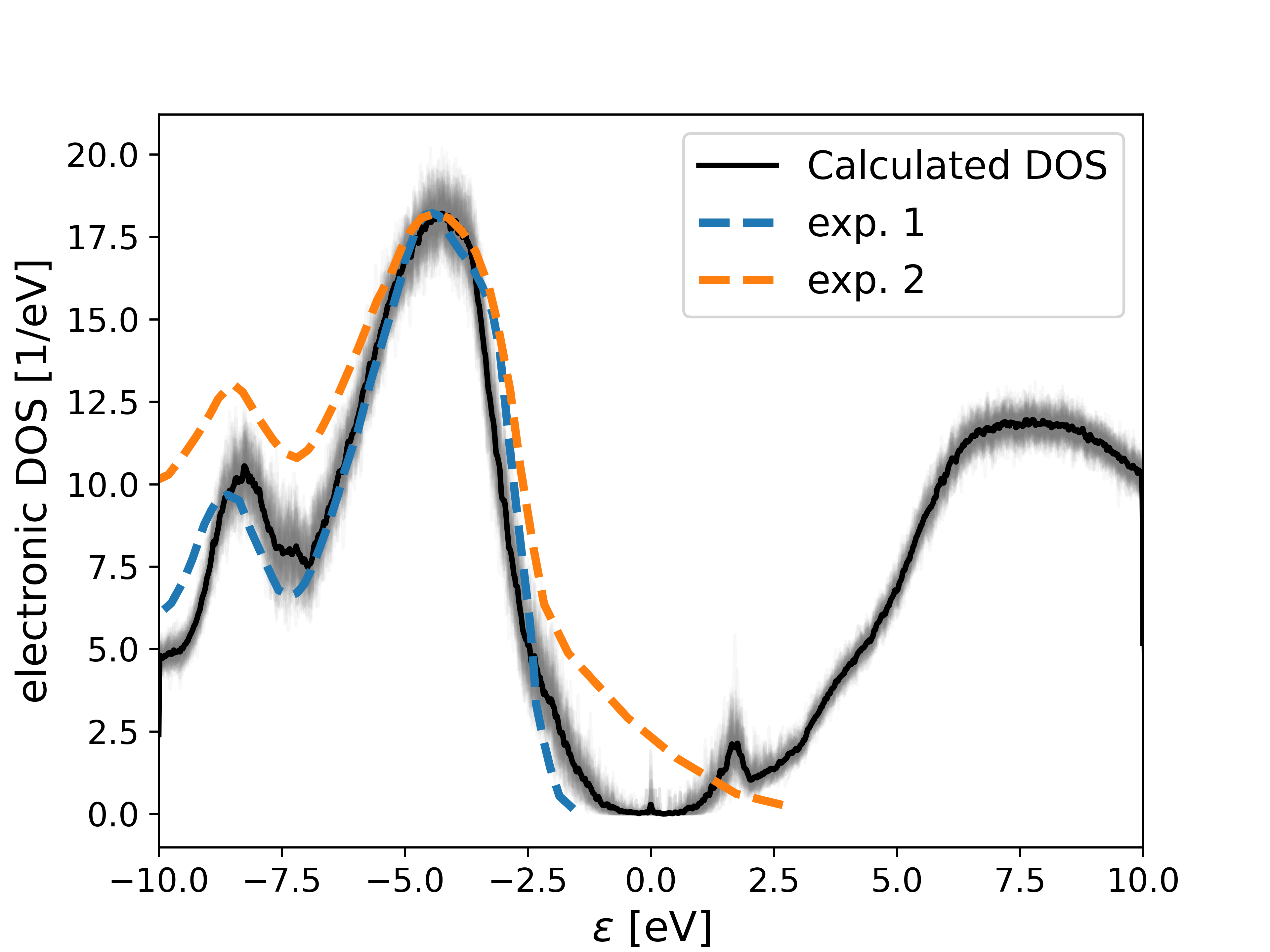}
\caption{\label{fig:el_dos} The PBE computed electronic density of states (DOS) averaged over the entire set of 3,000 random structures (solid black curve) is shown and compared against the photoemission (UPS) data from Refs.~\onlinecite{ups_exp1,ups_exp2} (dashed lines). The grey region shows average electronic DOSs of  the 100 random samples of 100 structures each.}
\end{figure}
%

Fig.~\ref{fig:el_dos} shows a comparison of the electronic DOS averaged over the entire set of 3,000 random structures (aligned on the common Fermi energy) and the photoemission (UPS) data from Refs.~\onlinecite{ups_exp1,ups_exp2}. The shaded grey region shows the average electronic DOSs of  the 100 random samples of 100 structures each. First, the overall good agreement between the computed DOS (black curve) and experiments validates both our approach, but also the physical picture it implies together with the procedure to align electronic DOSs of individual structures. Second, similar to the structure factor result,  the relatively narrow grey region around the black curve implies that, depending on the desired accuracy, as few as 100 random structures with 24 atoms (or possibly even less) could produce a satisfactory representation of the electronic structure of glassy SiO$_2$. This result further supports previous discussion and provides a route to modeling electronic properties of disordered systems, glasses but also other types of disorder, using high-accuracy electronic structure techniques.  One could even go a step further and formulate an approach in which DFT is used to obtain a fully converged results and then a subset of structures that best reproduce a fully converged result is selected for further calculations.
%
\subsection{Elastic properties -- bulk modulus}\label{ssec:bulk_mod}
%
%
\begin{figure}
\includegraphics[width=\linewidth]{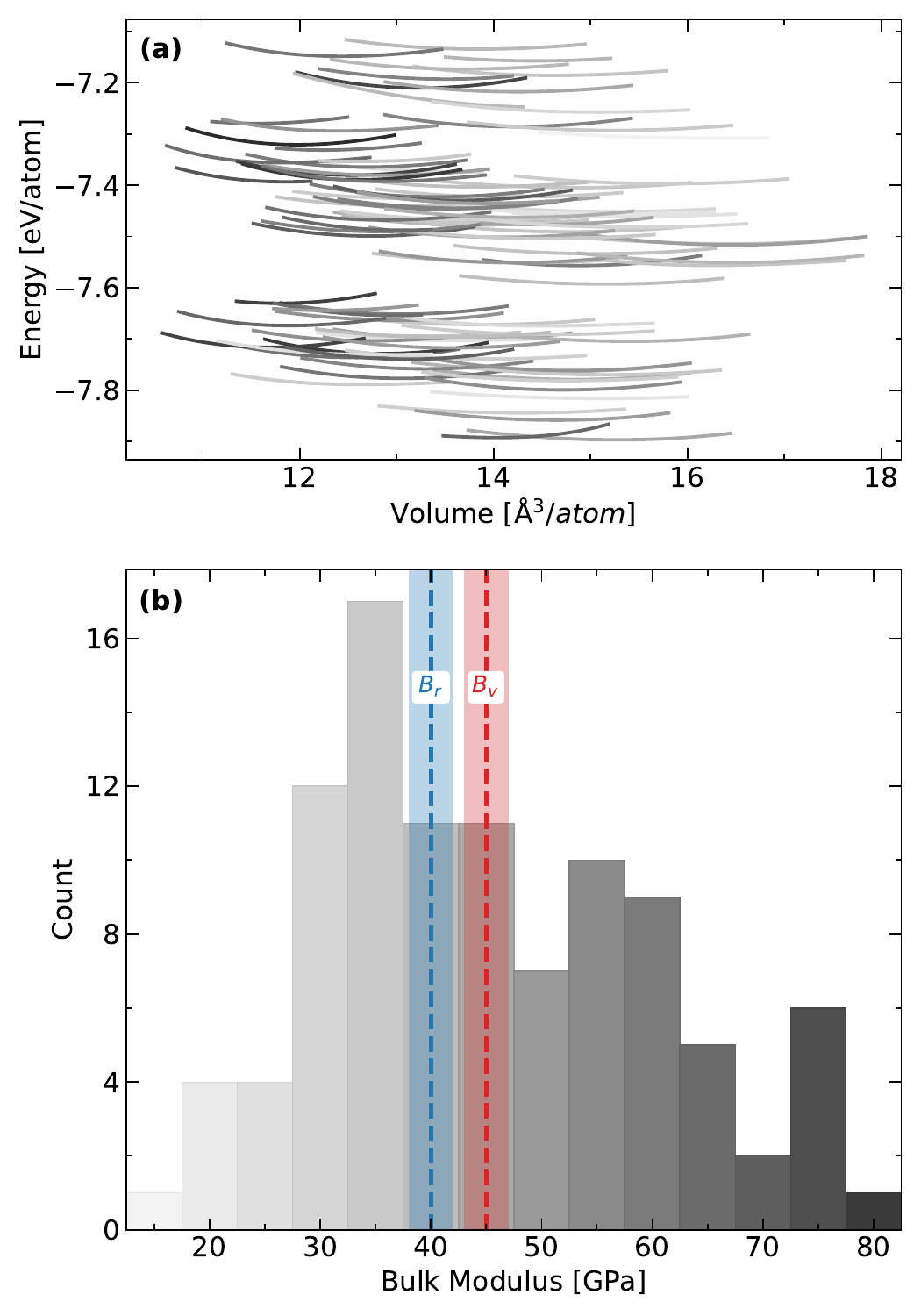}
\caption{\label{fig:bmod_histogram} (a) Equations of state for 100 randomly chosen structures from the set of 3,000 used in this work. (b) Histogram of the computed bulk modulus values with values and uncertainty of the Voigt and Reuss averages shown in blue and red respectively. Shades of grey show increasing value of the bulk modulus.}
\end{figure}
%
We further test the property calculations using a relatively small subset of the random structures by computing the bulk modulus of glassy SiO$_2$. For this purpose we choose randomly a subset of 100 structures (just one sample). For each structure from the sample, the bulk modulus value is obtained by fitting the Birch-Murnaghan equation of state \cite{Birch_PR:1947} to the discrete set of volume energy pairs. To do so, 6 scaled versions of the structures are created with scaling factors between $0.97$ and $1.03$. The scaled structures are then relaxed using PBE functional by relaxing all degrees of freedom except their volumes. Unlike the electronic DOS, the elastic properties cannot be directly averaged. To obtain the average value of the bulk modulus we again employ the picture of a glassy state as a composite of local environments. The Reuss and Voigt averaging schemes for the bulk modulus values of polycrystalline aggregates \cite{Bv_Br_def} are then used as the lower and upper bounds for the bulk modulus, respectively. They are computed in the following way:
\begin{equation}\label{eq:bulk_mod}
    B_{R} \, = \, \left(\sum_{i=1}^{n} \, \frac{V_{i}}{V \, B_{i}}\right)^{-1} \, \leq \, B \, \leq \, \sum_{i=1}^{M} \, \frac{V_{i}B_{i}}{V} \, = \, B_{V},
\end{equation}
where $V_{i}$ and $B_{i}$ are the volume and bulk modulus of the structure $i$, $V$ is the total volume of the system (sum of the volumes of structures), and $B$ is the actual bulk modulus of the system, both sums run over the number of considered structures $n$, we note that in this section $n$ will be used as the variable for the number of structures in a sample different to its previous definition as the global concentration. The lower and upper bounds are designated by $B_{R}$ and $B_{V}$, respectively. In order to estimate the uncertainty for the obtained values we use the central limit theorem, as has been done for alloy ensembles\cite{Novick_2023}. In order to apply the central limit theorem, both bounds need to be rewritten in terms of $B_{R} = \left<V\right> \, \times \, \left<\,\frac{V}{B} \, \right>^{-1}$ and $B_{V} = \frac{\left<VB\right>}{\left< V\right>}$ with
\begin{align}
    \left< V\right> &= \sum_{i=1}^{n}\frac{V_{i}}{n}\pm\frac{\sigma(V)}{\sqrt{n}} \nonumber\\ 
    \left< VB \right> &= \sum_{i=1}^{n} \frac{V_{i}B_{i}}{n}\pm\frac{\sigma(VB)}{\sqrt{n}} \\
    \left<\frac{V}{B}\right> &= \sum_{i=1}^{n}\frac{V_{i}}{B_{i}n}\pm\frac{\sigma(\frac{V}{B})}{\sqrt{n}},\nonumber
\end{align}
where $\sigma()$ represents the standard deviation of the quantity shown in parentheses, and $n$ is the number of structures in the sample. Following these equations, the sample of 100 randomly chosen structures results in  the bulk modulus bounds of $40 \pm 2\text{ GPa} \leq B \leq 45 \pm 2\text{ GPa}$.  Fits of Birch-Murnaghan equation of state for all 100 structure as well as a histogram of the bulk modulus values are shown in Fig.~\ref{fig:bmod_histogram}. The computed $B_{R}$ and $B_{V}$ values as well as their uncertainty are also shown as vertical blue and red dotted lines respectively. For comparison, different measured values of the bulk modulus of silica glasses are 36.8 GPa\cite{bmod_exp1},  38.6 GPa\cite{bmod_exp2} or 37 GPa\cite{bmod_exp3}. Knowing that typical errors DFT methods make when evaluating bulk modules of crystalline solids is $\sim$10 \% we conclude that the calculated bulk modulus of glassy SiO$_2$ obtained using our representation of the glassy state is in a good agreement with measurements. 

\section{Conclusions}
%
In conclusion, we present in this paper the first-principles approach to modeling glasses as a collection (composite) of small-cell, periodic local minima on the potential energy surface. The local minima are obtained using the random structure sampling, that is, by generating a relatively large set of random structures which are subsequently relaxed to the closest local minimum using first-principles methods.  Within this approach the static properties of glasses than become averages over the entire set of the random structures. Using glassy SiO$_2$ as an example we show how the (average) properties evaluated in our model agree well with the measured ones. These properties include mass density, structure factors, pair distribution functions, as well as the electronic density of states and  the bulk modulus. The significance of the method presented here is first in its truly first-principles nature, as no experimental parameters is required, but most importanty this method also offers a new computationally tractable route for predicting functional properties of glasses, most notably optical ones, that are often unatainable to more traditional methods.
%

\section*{Acknowledgements}
This work is supported by the National Science Foundation, Grant No. DMR-1945010 and was performed using computational resources of Colorado School of Mines. A. Novick acknowledges the support of NSF Grant No.OAC-2118201. 

\section*{Data Availability}
The data that support the findings of this study are available from the corresponding author upon reasonable request.

\appendix

\section{Aschroft-Langreth formalism for multicomponent systems}

\noindent Within the Aschroft-Langreth (AL) formalism \cite{Ashcroft_1967} the total structure factor is related to the coherent scattering intensity per atom as:
\begin{equation}
S^{\mathrm{AL}}({\bf q})) \, = \, \frac{I({\bf q})}{\langle f({\bf q})^2\rangle},
\end{equation}
where $\langle f({\bf q}))^2\rangle$ is the average squared atomic form factor.
\noindent Partial Aschroft-Langreth structure factors for multicomponent systems are defined as:
\begin{equation}
S_{\alpha\beta}^{\mathrm{AL}} ({\bf q})\, = \, \frac{1}{\sqrt{N_{\alpha}N_{\beta}}} \, \sum_{j \in \alpha}^{N_{\alpha}}\sum_{k \in \beta}^{N_{\beta}} \, e^{-i{\bf q}{\bf r}_{jk}}  -  \sqrt{N_{\alpha}N_{\beta}} \, \Delta_{{\bf q}0}.
\end{equation}
The second term having the Kronecker delta $\Delta_{{\bf q}0}$ serves to remove the forward scattering (${\bf q}=0$) contribution. The total Aschroft-Langreth structure factor can then be expressed as the sum of the partials as:
\begin{equation}\label{eq:tot_s_AL}
S^{\mathrm{AL}} ({\bf q})\, = \,  \sum_{\alpha}\sum_{\beta} \, \sqrt{c_{\alpha}c_{\beta}} \, \frac{f_{\alpha}({q})f_{\beta}({q})}{\langle f({q})^2 \rangle} \,  S_{\alpha\beta}^{\mathrm{AL}}({\bf q}).
\end{equation}
With partial pair distribution functions that are defined as:
\begin{equation}\label{eq:f_def}
g_{\alpha\beta} \, = \, \frac{1}{n c_{\beta}} \, \left[ \frac{1}{N_{\alpha}} \, \sum_{j \in \alpha} \sum_{k \in \beta} \, \delta({\bf r} - {\bf r}_{jk} ) - \Delta_{\alpha\beta}\delta({\bf r})  \right],
\end{equation}
one can obtain the following expression by applying $\int [\dots] e^{-i{\bf q \, r}} \, d{\bf r}$ on the entire previous equation (integration goes over the volume occupied by the system):
\begin{equation}
S_{\alpha\beta}^{\mathrm{AL}} ({\bf q}) \, = \, \Delta_{\alpha\beta} + n \sqrt{c_{\alpha}c_{\beta}} \int [g_{\alpha\beta}({\bf r}) -1] \, e^{-i{\bf qr}} d{\bf r}.
\end{equation}
In the above derivation the following definition of the Kronecker delta is used $\Delta_{{\bf q}0} = 1/V \, \int \, e^{-i{\bf q \, r}} \, d{\bf r}$, where it is assumed that either the volume $V$ of the system consists of the integer number of periods of the plane wave $e^{-i{\bf q \, r}}$ for any ${\bf q}$, or that the $V$ is large enough so that any nonzero value of the integral for ${\bf q} \neq 0$ when divided by $V$ is infinitesimally small (zero for practical purposes). For homogeneous and isotropic systems $g_{\alpha\beta}({\bf r})=g_{\alpha\beta}(r)$ and the integration over the angular degrees of freedom can be carried out resulting in:
\begin{equation}
S_{\alpha\beta}^{\mathrm{AL}} (q) \, = \, \Delta_{\alpha\beta} + n \sqrt{c_{\alpha}c_{\beta}} \int 4\pi r [g_{\alpha\beta}(r) -1] \, \frac{sin(qr)}{q} dr.
\end{equation}
Note that the dependence of the PDF on $r=|{\bf r}|$ automatically implies dependence of $S_{\alpha\beta}^{\mathrm{AL}}$ on $q=|{\bf q}|$. By introducing a reduced partial PDF as $D_{\alpha\beta}(r) = n4\pi r [g_{\alpha\beta}(r) -1]$ one can write:
\begin{equation}
S_{\alpha\beta}^{\mathrm{AL}} (q) \, = \, \Delta_{\alpha\beta} + \sqrt{c_{\alpha}c_{\beta}} \int D_{\alpha\beta}(r) \, \frac{sin(qr)}{q} dr.
\end{equation}
Conversely, by performing $2/\pi \, \int [\dots] sin(qr') \, dq$ on the previous equation one gets:
\begin{equation}
D_{\alpha\beta}(r') \, = \, \frac{2}{\pi} \frac{1}{\sqrt{c_{\alpha}c_{\beta}}} \int_{0}^{+\infty} \, q \, [S_{\alpha\beta}^{\mathrm{AL}} (q) - \Delta_{\alpha\beta}] \, sin(qr') \, dq
\end{equation}
This equation establishes a relationship between the reciprocal space and real space properties. Often, a relationship similar to this is used having the total structure factor under the integral instead of the partial ones. In that case one gets:
\begin{equation}
\begin{gathered}
\frac{2}{\pi} \, \int_0^{+\infty} \, q \, [ S^{\mathrm{AL}} (q) - 1 ] \, sin(qr) \, dq  = \\
= \sum_{\alpha}\sum_{\beta} \sqrt{c_{\alpha}c_{\beta}} \frac{2}{\pi}  \int_0^{+\infty} \frac{f_{\alpha}(q)f_{\beta}(q)}{\langle f(q)^2 \rangle} \,\times \\
\times \, q \, [ S_{\alpha\beta}^{\mathrm{AL}} (q) - \Delta_{\alpha\beta}] \, sin(qr) \, dq \, = \\
= \sum_{\alpha}\sum_{\beta} c_{\alpha}c_{\beta} \frac{Z_{\alpha}Z_{\beta}}{\langle Z^2\rangle} D_{\alpha\beta} (r) = D(r).
\end{gathered}
\end{equation}
In this way a total reduced PDF $D(r)$ is obtained. In the derivation the Warren \textit{et al.} approximation $f_{\alpha}(q) = Z_{\alpha} \, h(q)$ is used, with $h(q)$ being approximately universal, atom-independent function of $q$ \cite{Warren_1936} as already discussed. It is important to note, that strictly speaking $D(r)$ is not directly connected with the total PDF $g(r)$. This is only the case for single component systems for which   $Z_{\alpha}Z_{\beta}/\langle Z \rangle^2 = 1$ where $g(r) = 1+1/n4\pi r \, D(r)$.

\section{Faber-Ziman formalism for multicomponent systems}
%
In the Faber-Ziman formulation the partial structure factors can be expressed as:
\begin{equation}\label{eq:part_FZ}
S_{\alpha\beta}^{\mathrm{FZ}} ({\bf q}) \, = \,  1 + n \int [g_{\alpha\beta}({\bf r}) -1] \, e^{-i{\bf qr}} d{\bf r}.
\end{equation}
As before, in case of isotropic systems this expression becomes:
\begin{equation}
S_{\alpha\beta}^{\mathrm{FZ}} (q) \, = \,  1 + n \int 4\pi r \, [g_{\alpha\beta}(r) -1] \, \frac{sin(qr)}{q} dr,
\end{equation}
and the inverse:
\begin{equation}
g_{\alpha\beta}(r) = 1 + \frac{1}{n 4\pi r} \frac{2}{\pi} \int_{0}^{+\infty} q \, [S_{\alpha\beta}^{\mathrm{FZ}} (q) -  1] sin(qr) dq.
\end{equation}

\noindent The relationship between the Faber-Ziman and Aschroft-Langreth partial structure factors can be obtained by substituting the expression for $g_{\alpha\beta}({\bf r})$ from Eq.~(\ref{eq:f_def}) in Eq.~(\ref{eq:part_FZ}) produces:
\begin{equation}
S_{\alpha\beta}^{\mathrm{FZ}} ({\bf q}) \, = \, \frac{1}{\sqrt{c_{\alpha}c_{\beta}}}  S_{\alpha\beta}^{\mathrm{AL}} ({\bf q}) - \frac{1}{c_{\beta}}\Delta_{\alpha\beta} + 1.
\end{equation}
In this formalism the total structure factor is:
\begin{equation}
S^{\mathrm{FZ}} ({\bf q}) = \frac{1}{\langle f(q) \rangle^2} I({\bf q}) - \frac{\langle f(q)^2 \rangle - \langle f(q) \rangle^2}{\langle f(q) \rangle^2},
\end{equation}
or in terms of the Faber-Ziman partial structure factors:
\begin{equation}
S^{\mathrm{FZ}} ({\bf q}) = \sum_{\alpha}\sum_{\beta} \, c_{\alpha}c_{\beta} \frac{f_{\alpha}(q) f_{\beta}(q)}{\langle f(q) \rangle^2} S_{\alpha\beta}^{\mathrm{FZ}} ({\bf q}).
\end{equation}
Lastly, the Faber-Ziman total structure factor expressed in terms of the Aschroft-Langreth partial structure factors $S_{\alpha\beta}^{\mathrm{AL}} ({\bf q})$ becomes:
\begin{equation}
S^{\mathrm{FZ}} ({\bf q}) = \sum_{\alpha}\sum_{\beta} \sqrt{c_{\alpha}c_{\beta}}  \frac{f_{\alpha}(q) f_{\beta}(q)}{\langle f(q) \rangle^2} \left[ S_{\alpha\beta}^{\mathrm{AL}} ({\bf q}) - \Delta_{\alpha\beta} \right] + 1.
\end{equation}
This last expression is used throughout our work as the experimentally reported structure factors are often provided in the Faber-Ziman way.
\bibliography{biblio.bib} 

\end{document}